# Evidence for coherent spicule oscillations by correcting *Hinode*/SOT Ca II H in the southeast limb of the Sun

A. R. Ahangarzadeh Maralani[1]★, E. Tavabi[2]★ and A. Ajabshirizadeh[3]

[1] *Department of Physics, Tabriz Branch, Islamic Azad University, Tabriz, Iran*
[2] *Physics Department, Payame Noor University (PNU), 19395-3697-Tehran, I. R. of Iran*
[3] *Department of Theoretical Physics and Astrophysics, University of Tabriz, 51664, Tabriz*

**ABSTRACT**

Wave theories of heating the chromosphere, corona, and solar wind due to photospheric fluctuations are strengthened by the existence of observed wave coherency up to the transition region (TR). The coherency of solar spicules' intensity oscillations was explored using the *Solar Optical Telescope* (SOT) on the *Hinode* spacecraft with a height increase above the solar limb in active region (AR). We used time sequences near the southeast region from the *Hinode*/SOT in Ca II H line obtained on April 3, 2015 and applied the de-convolution procedure to the spicule in order to illustrate how effectively our restoration method works on fine structures such as spicules. Moreover, the intensity oscillations at different heights above the solar limb were analysed through wavelet transforms. Afterwards, the phase difference was measured among oscillations at two certain heights in search of evidence for coherent oscillations. The results of wavelet transformations revealed dominant period peaks in 2, 4, 5.5, and 6.5 min at four separate heights. The dominant frequencies for coherency level higher than 75 percent was found to be around 5.5 and 8.5 mHz. Mean phase speeds of 155-360 km s$^{-1}$ were measured. We found that the mean phase speeds increased with height. The results suggest that the energy flux carried by coherent waves into the corona and heliosphere may be several times larger than previous estimates that were based solely on constant velocities. We provide compelling evidence for the existence of upwardly propagating coherent waves.

**Keywords:** Sun: oscillations – Sun: chromosphere – Sun: transition region – methods: observational – techniques: image processing.

## 1 INTRODUCTION

The interface between the lower solar atmosphere, where energy is generated by subsurface convection and the corona, comprises the chromosphere which is dominated by jet-like, dynamic structures. These are called mottles when found in quiet regions, fibrils when initiated in active regions, and spicules when observed at a solar limb. Space observations with *Hinode* have two different types of spicules, namely type-I and type-II, each having different properties (De Pontieu et al. 2007; Tsiropoula et al. 2012).

Pereira, De Pontieu, and Carlsson (2012) found clear evidence for the existence of two types of spicules. Type-I spicules revealed a rise and fall and had typical lifetimes of 150-400 s and maximum ascending velocities of 15-40 km s$^{-1}$, while type-II spicules enjoyed shorter lifetimes of 50-150 s, much faster velocities of 30-110 km s$^{-1}$, and were not seen to fall down in the Ca II H filter. Rather, they tend to fade at around their maximum length. Type-II spicules are found to be the most common and were seen in quiet sun and coronal holes. Type-I spicules have been mostly spotted in active regions. The fading of type-II spicules had reinforced the idea that they were rapidly heated to higher temperatures, not visible in the Ca II H filter, and might deposit mass and energy at coronal heights (De Pontieu et al. 2007, 2009).

Madjarska, Vanninathan, and Doyle (2011) analysed three large spicules and found that they comprised numerous thin spicules which rose, rotated, and descended simultaneously forming a bush-like feature. Their rotation resembled the untwisting of a large flux rope. They showed velocities ranging from 50-250 km s$^{-1}$.

Martinez-Sykora et al. (2011) demonstrated that the combination of strong Lorentz forces in the upper chromosphere could produce high-velocity jets of cool material along with the magnetic field of the corona. These jets share several characteristics with the observed type-II spicules. Goodman (2012) used a time dependent magneto-hydrodynamic model. He confirmed that type-II spicule velocities could be generated by a Lorentz force under chromospheric conditions, and maximum vertical flow could reach a speed of 150-460 km s$^{-1}$.

Gupta et al. (2013) detected period bands of 2-4 min, 4-6 min, 6-15 min, and long–period oscillations with periods between 15 and 30 min in bright magnetic points from the chromosphere to the transition region. He et al. (2009) reported that spicules were modulated by relatively high frequency (>0.02 Hz) transverse fluctuations that propagated upward along with the spicules with phase speed ranges from 50-150 km s$^{-1}$. Some of these modulated spicules displayed clear wave-like shapes with wavelengths shorter than 8 Mm.

Using images with high temporal and spatial resolutions obtained from *Hinode*/SOT, De Pontieu et al. (2007) discovered that the chromosphere (i.e., the region between the solar surface and the corona) are permeated by Alfvén waves with amplitudes on the order of 10 to 25 km s$^{-1}$ and periods of 100 to 500 s. The energy fluxes carried by these waves are estimated to be energetic enough to accelerate the solar wind and possibly to heat the quiet corona.

Tavabi (2014) investigated off-limb and on-disk spicules to detect a counterpart of the limb spicule on the disk and found a definite signature with strong power in 3 min (5.5 mHz) and 5 min (3.5 mHz), a full range of oscillations, and a high frequency of intensity fluctuation (greater than 10 mHz or less than 100 s) corresponding to the occurrence of type-II spicules.

Using spectral imaging data in the Ca II 854.2 nm and Hα lines with *Crisp Imaging Spectro Polarimeter* (CRISP) at the Swedish Solar Telescope (SST) in La Palma, Rouppe van der Voort et al. (2009) reported the discovery of disk counterparts for type-II spicule. They showed that the attributes of these rapid blue excursions on the disk are very similar to those of type-II spicules at the limb. A detailed study of the spectral line profiles in these events suggests that plasma is accelerated along with the jet and is heated throughout the short lifetime of the event.

Using observations from the *Interface Region Imaging Spectrograph* (IRIS) through direct imaging on the solar disk,

★Email: ahangarzadeh@iaut.ac.ir, tavabi@iap.fr

Tian et al. (2014) reported that high-speed-up flows appear at speeds of 80-250 km s$^{-1}$.

De Pontieu et al. (2009, 2011) came to this finding that type-II spicules represented an impulsively accelerated chromospheric material that is continuously heated while rising. The causes of the heating and acceleration are unknown. However, a magnetic process such as reconnection is likely to be important.

De Pontieu et al. (2012) used high-quality observations with SST to establish that type-II spicules are characterized by three different types of motion, namely field-aligned flows of order 50-100 km s$^{-1}$, swaying motions of order 15-20 km s$^{-1}$, and torsional motions of order 25-30 km s$^{-1}$. Their analysis yielded strong evidence that most, if not all, type-II spicules undergo large torsional modulations. In addition, these motions, like spicule swaying, represent Alfvenic waves propagating outward at several hundred km s$^{-1}$. Moreover, they found that large torsional motion was an ingredient in the production of type-II spicules and spicules played an important role in the transfer of helicity through the solar atmosphere.

Through making coordinated observations with IRIS and SST, Rouppe van der Voort et al. (2015) found spectral signatures for rapid blue- and red-shifted excursions in C II 1335 Å and 1336 Å and Si IV 1394 Å and 1403 Å spectral lines. They interpreted this as a clue that type-II spicules are heated to at least TR temperatures.

Tavabi et al. (2015b) found that surge-like behavior of solar polar region spicules supported the untwisting multi-componental interpretation of spicules exhibiting helical dynamics. They analysed the proper transverse motions of mature and tall polar region spicules at different heights in detail. Assuming that there might exist Helical-Kink waves or Alfvenic waves propagating inside their multi-componental substructure, they interpreted the quasi-coherent behavior of all visible components presumably confined by a surrounding magnetic envelope. In addition, they concentrated their analysis on the taller Ca II spicules which is more pertinent to coronal heights and easy to measure.

The observed contrast of small scale features like spicules is influenced by the point spread function (PSF). In particular, the scattered light strongly reduces the contrast. Spicules, which are small structures, are strongly affected (Mathew, Zakharov, & Solanki, 2009). Optical devices such as telescopes could not work as effectively as their diffraction limited quality due to optical aberration arising from imperfections, thermal changes and impurities. They diverge from optimum adjustment owing to the passage of time. Even a small dust or scratch may produce a significantly scattered light. Scattered light is mathematically described by using the extended wings of the PSF, which causes an image of the point source of light to be largely spread (Richardson, 1972; Lucy, 1974).

In the present study, we used time sequences taken by the *Hinode*/SOT in Ca II H (396.8 nm) line. We studied the frequency of spicules' oscillations and their coherency at different heights.

## 2 OBSERVATIONS

We used time sequences of AR data taken near the southeast region of solar limb observed in the Ca II H line with the *Hinode*/SOT (the wavelength pass-band was centred at 396.8 nm with a FWHM of 0.3 nm). The characteristics of the observation series are summarized in Table 1.

**Table 1** The dataset obtained with the *Hinode*/SOT in the Ca II H line.

| Date | Start and end times (U.T.) | X-center Y-center (arcsec) | Cadence (sec.) | Size (Pixels$^2$) | X-FOV Y-FOV (arcsec) | Pixel size (arcsec) |
|---|---|---|---|---|---|---|
| 2015 April 3 | 00:39:30 01:26:38 | -848.1 -497.9 | 5 | 512×512 | 55.787 55.787 | 0.109 ~80km |

The standard SOT subroutines (WV_DENOISE, FG_SHIFT_PIX, UNSHARP_MASK, FG_PREP, and FG_BAD_PIX) were used to reduce the image spikes, jitter, align the time series, calibrate raw data, correct the CCD readout anomalies, bad pixels, and flat-fields, and subtract the dark pedestal.

## 3 DATA ANALYSIS

In order to reduce scattered and/or stray light in the extended wings of the PSF (Mathew, Zakarov, & Solanki 2009), the Max-likelihood routine was applied to all images. Due to the use of IDL Max-likelihood program, Root Mean Square (RMS) contrast of the images went up. The normalized RMS contrasts are factors of 1.61 higher than those obtained before PSF de-convolution.

Then, Mad-max operator was applied to improve the visibility of fine features and to clarify the images (see Fig. 1d). The operator considerably reduced background noise (Koutchmy & Koutchmy, 1989; November & Koutchmy, 1996; Tavabi, Koutchmy, & Ajabshirizadeh, 2013; Tavabi 2014).

Spicules displayed radial outflow, suggesting Alfvenic wave propagation at about 300 km s$^{−1}$ outward along with the spicules (De Pontieu et al. 2007b, 2012). The continuation of these waves into the transition region and corona indicates that they can help to drive away the solar wind (McIntosh et al. 2011).

Most spicules oriented approximately perpendicular to the surface of the Sun in the radial direction (see Fig. 1d). Therefore, the fluctuations were analysed at different heights above the solar limb in X and Y axes, simultaneously. We measured phase speeds in radial direction. The sum of the two phase speeds along the X and Y axes was approximately equal to the phase speed along the spicules. To this end, diagrams of time-slices were prepared using the highly processed images to look for spatial variations of the spicules with temporal changes.

An algorithm for the time-X/Y cut diagram puts together a specific row/column of all processed images from top to bottom and from left to right. We analysed all pixels in the X and Y directions. Fig. 2 depicts only one example for Y = −517 arcsec, and Fig. 3 shows for X= -854 arcsec (Tavabi 2014).

The wavelet results obtained in four separate heights are marked by 1, 2, 3, and 4 red color lines in Fig. 2 and Fig. 3. The heights start from clinging to the solar limb to out of the limb equivalent to 4200 km (at a distance of 1400 km from each other, respectively). The wavelet results showed dominant periods about 2, 4, 5.5, and 6.5 min (see Fig. 4 for time-X cut series and Fig. 5 for time-Y cut series).

Detailed wavelet analysis which supply information on the temporal variation of a signal are described in Torrence and Compo (1998). The Morlet function was used in line with the definition given in Torrence and Compo (1998) and Ajabshirizadeh, Tavabi and Koutchmy (2008) for convolution with time series in wavelet transformation.

Similar oscillations in all time sequences at four separate heights were identified by comparing the results of wavelet analysis for time-X cut, and time-Y cut series.

We used cross-spectrum phase, coherence estimates, and the cross-wavelet transform software package (Torrence & Compo, 1998) between two heights and calculated the coherency, phase differences, and cross-wavelet oscillations (Fig. 6, and Fig. 7, for time-X cut and time-Y cut series, respectively; see 2D representation in Tavabi et al. 2015a; 2D and 3D representation in Zeighami et al. 2016). The findings of the coherent oscillations between two different heights showed coherency in periods of about 2 and 3 min.

Measurements of the phase difference between two intensity profiles obtained at the two heights yielded information about the phase delay between oscillations as a function of frequency (Trauth, 2007; Oppenheim & Verghese, 2010). These calculations were made for maximum coherency higher than 75 %. The results of the phase difference analysis revealed coherency at frequencies similar to those of the cross-wavelet results (Fig. 8 panels a, and b, for time-X cut and time-Y cut series, respectively).

Using the equation for phase difference as a function of frequency, $\Delta \varphi = 2\pi f T$ where $f$ is the frequency (in Hz) and $T$ is the time difference (in seconds), we obtained the time delay between two adjacent layers. Afterwards, we calculated the phase speed between these heights using the equation $V_{ph} = H/T$, where $H$ is the distance between the two heights. We computed the phase difference and phase speed for maximum coherency from all the coherency plots. Through this procedure, we took all the time-slice results into consideration. The mean phase speeds between two heights were 85, 140, and 195 km s$^{-1}$ for time-X cut series and 130, 220, and 305 km s$^{-1}$ correspondingly for time-Y cut series, respectively. We found that the mean phase speed increased with height.

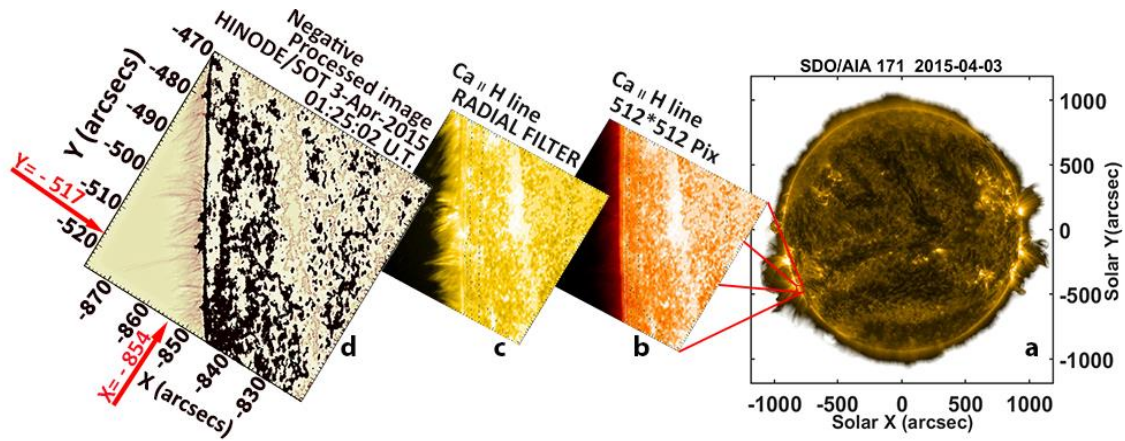

**Figure 1.** (a) A full-disk image on April 3, 2015, obtained with the Atmospheric Imaging Assembly (AIA) onboard the Solar Dynamics Observatory (SDO). The AR is clearly visible. (b) An image obtained with *Hinode*/SOT in Ca II H line (396.8 nm), and (c) an image in Ca II H line with Radial filter on April 3, 2015. (d) A highly processed negative image sample to improve the visibility of fine features after applying the mad-max and Max-likelihood operators obtained with *Hinode*/SOT.

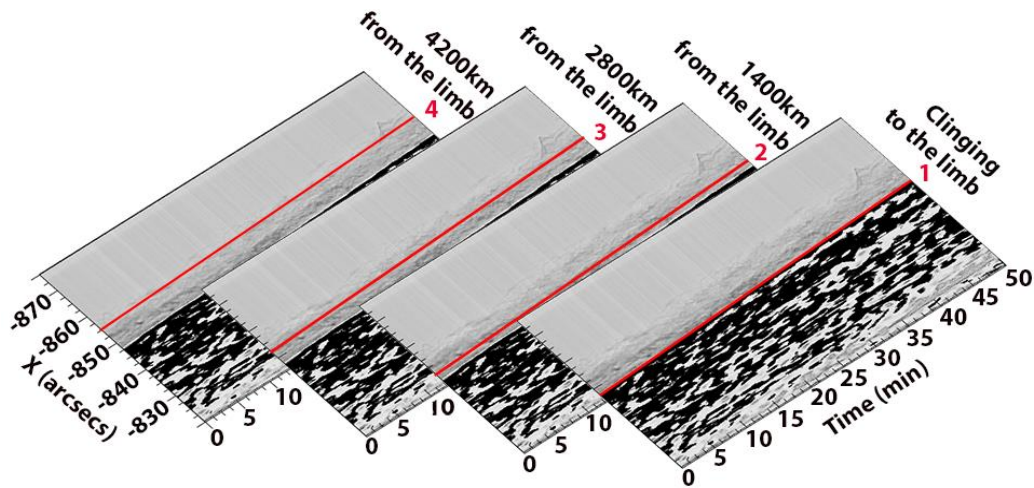

**Figure 2.** Time-X cut diagrams in negative for Y=-517 arcsec shown in Fig. 1d. The positions of four separate heights which were selected for analysis are marked by four red color lines (1, 2, 3, and 4) at a distance 1400 km from each other. A drift speed toward the east was identified from solar limb motion. The drift had an average speed of lower than 0.064"/min.

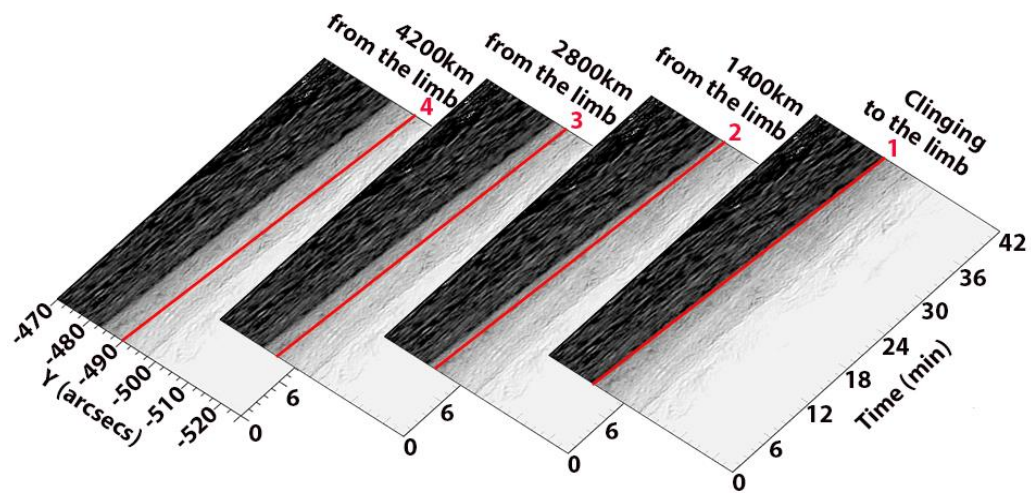

**Figure 3.** Time-Y cut diagrams in negative for X= -854 arcsec shown in Fig. 1d. A drift speed toward the south was identified from solar limb motion. The drift had an average speed of lower than 0.092"/min. The red color lines have a similar meaning as in Fig.2.

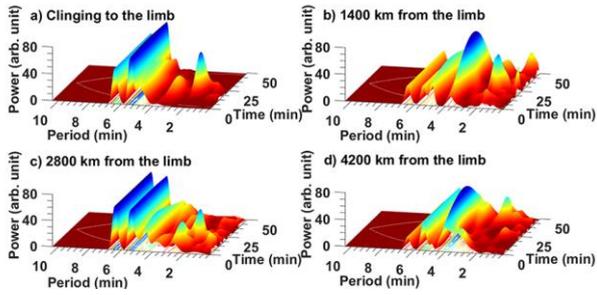

**Figure 4.** (a)-(d) represent the 3-D wavelet results obtained from intensity profiles for time-X cut series corresponding to the four heights which are displayed by four red color lines in Fig. 2 (for Y= - 517 arcsec). The solid white curve in the plot at each panel indicates the cone of influence region (COI) where the wavelet power spectra are distorted because of the influence of the end points of finite length signals.

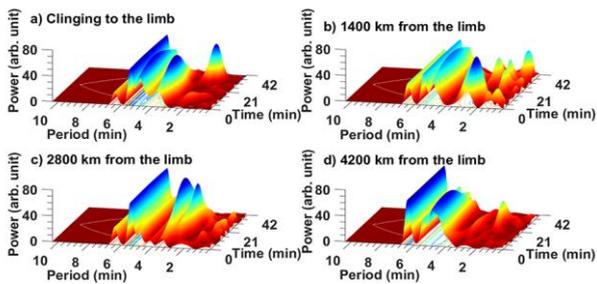

**Figure 5**. (a)-(d) represent an example of the 3-D wavelet results obtained from intensity profiles for time-Y cut series corresponding to the four heights that are marked by four red color lines in Fig. 3 (corresponding to X= - 854 arcsec). The white curve in the plot at each panel has a similar meaning as in Fig. 4.

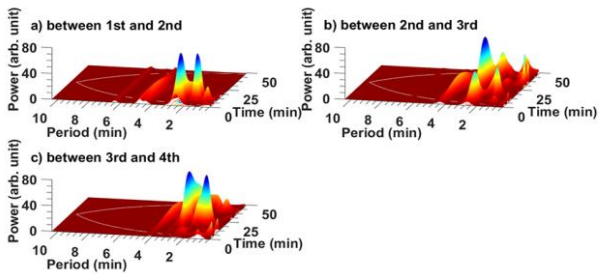

**Figure 6.** An example of 3-D wavelet power-spectrum results for coherency between two heights for time-X cut series. (a)-(c) show the wavelet coherency spectrum between two heights, i.e., 1$^{st}$ and 2$^{nd}$, 2$^{nd}$ and 3$^{rd}$, 3$^{rd}$ and 4$^{th}$ heights, respectively (for Y= - 517 arcsec). The white curve in the plot at each panel has a similar meaning as in Figs. 4 and 5.

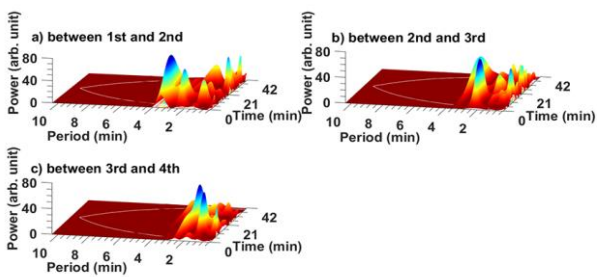

**Figure 7.** An example of 3-D wavelet power-spectrum results for coherency between two heights for time-Y cut series. (a)-(c) show the wavelet coherency spectrum between the two heights, i.e., 1$^{st}$ and 2$^{nd}$, 2$^{nd}$ and 3$^{rd}$, 3$^{rd}$ and 4$^{th}$ heights, respectively (for X= - 854 arcsec). The white curve in the plot at each panel has a similar meaning as in Figs. 4, 5, and 6.

## 4 DISCUSSION AND CONCLUSION

The findings revealed that the use of the IDL Max-likelihood program led to an increase in the RMS contrast of the images. The normalized RMS contrasts are factors of 1.61 higher than those obtained before PSF de-convolution.

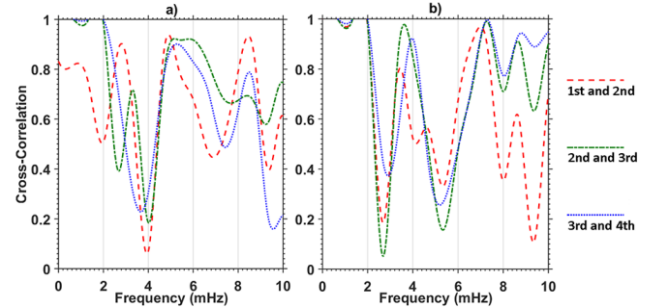

**Figure 8.** Examples of coherency plots. The plots show coherency of waves propagated between different heights. Panel's a plot for time-X cut series, and panel's b plot for time-Y cut series. Dashed lines are for coherency between 1$^{st}$ and 2$^{nd}$ heights, dotted dash lines for coherency between 2$^{nd}$ and 3$^{rd}$ heights, and dotted lines for coherency between 3$^{rd}$ and 4$^{th}$ heights.

The wavelet results indicated dominant periods of about 2 min, 4 min, 5.5 min, and 6.5 min, corresponding to the findings of Zaqarashvili and Erdélyi (2009), Gupta et al. (2013), Hansteen, Betta and Carlsson, (2000), and Tavabi (2014). As expected, the results of coherency at two certain heights represented periods at about 2, and 3 min oscillations. The dominant frequency for coherency level exceeding 75 percent showed frequencies of about 5.5, and 8.5 mHz. The mean phase speeds between the two heights were 85, 140, and 195 km s$^{-1}$, for time-X cut series and 130, 220, and 305 km s$^{-1}$ for time-Y cut series, respectively. The findings emanating from these two-speed components provided mean phase speeds of 155-360 km s$^{-1}$ for AR (McIntosh et al. 2011; Okamoto & De Pontieu 2011).

Tavabi et al. (2015a) used three-time sequences for the quiet Sun (22, November, 2006), active Sun (15, January, 2014), and active region (26, January, 2007) from the equator of the solar limb observed in Ca II H line by *Hinode*/SOT. The mean phase speeds measured by Tavabi et al. (2015a) for the quiet Sun were found to be 50-450 km s$^{-1}$, 50-650 km s$^{-1}$ for the active region, and 50-550 km s$^{-1}$ for the active Sun. Zeighami et al. (2016) used time sequences of AR data (26, January, 2007) taken at the solar equatorial limb observed in the Ca II H line with the *Hinode*/SOT. They reported mean phase speeds within the range of 250-425 km s$^{-1}$ for AR data.

Furthermore, we clearly observe that the phase speed increases with height. This could be caused by two parameters. For one thing, for greater heights, the Alfven speed is higher since the plasma density is low. Considering that the magnetic field strength is expected to decrease with height, the range of phase speeds at heights below 4200 km seems reasonable.

Similar results were obtained by Okamoto and De Pontieu (2011) for individual packets of transverse waves by analysing the horizontal displacement of the medial axes of the spicules as a function of time and height. However, as described in Okamoto and De Pontieu (2011), the phase speeds at heights above 15 arcsec (6000 km) appear to be too high to be explained in terms of the change in Alfven speed. Another reason is that the upward and downward propagating waves are more often superposed at greater heights. This would arise naturally in a scenario in which numerous waves are excited at lower heights, reflected at the TR at the top of spicules, thereby resulting in downward propagating waves at greater heights (just below the top). Okamoto and De Pontieu (2011) observed a combination of upward and downward propagating waves near the top of spicules. Regarding the high reflection coefficient of the TR (Hollweg, Jackson, & Galloway; 1982), it is expected that the downward and upward propagating waves have roughly equal amplitudes. However, since some of the wave energy leaks into the corona, the amplitudes are not equal.

Superposition of the counter propagating waves could lead to partially standing waves with very high phase speeds. To our best knowledge, the reflected waves have not been found to reach back down to the bottom of the spicules. The increase in Alfven speed and the reflection of the top of the spicule provide a good justification for the upsurge in phase speed in parallel with height.

A comparison of the findings obtained from this study with those of Tavabi et al. 2015a, Zeighami et al. 2016 indicated that dominant frequencies and mean phase speeds might be dependent on the location of spicules in the limb of the sun, solar cycle, and active region.

## ACKNOWLEDGMENTS

*Hinode* is a Japanese mission developed and launched by ISAS/JAXA, with NAOJ as its domestic partner and NASA and STFC (UK) as its international partners. The image processing program Mad-max was provided by O. Koutchmy, see http://www.ann.jussieu.fr/~koutchmy/debruitage/madmax.pro and the wavelet analysis software was provided by Torrence and Compo (http://atoc.Colorado.edu/research/wavelets). This study was financially supported by Islamic Azad University, Tabriz Branch (IAUT).